\documentclass[conference]{IEEEtran}
\IEEEoverridecommandlockouts
\usepackage{amsmath,amssymb,amsfonts}
\usepackage{algorithmic}
\usepackage{graphicx}
\usepackage{textcomp}
\usepackage{xcolor}
\usepackage{hyperref}
\def\BibTeX{{\rm B\kern-.05em{\sc i\kern-.025em b}\kern-.08em
    T\kern-.1667em\lower.7ex\hbox{E}\kern-.125emX}}
\usepackage{multirow}
\DeclareMathOperator*{\argmin}{arg\,min}
\def\x{{\mathbf x}}
\def\X{{\mathbf X}}
\def\S{{\mathbf S}}
\def\G{{\mathbf G}}
\def\I{{\mathbf I}}

\usepackage[eprint=false,isbn=false,url=false,style=ieee,maxnames=5,minnames=1]{biblatex}   
\AtEveryBibitem{% Clean up the bibtex rather than editing it
 \clearlist{address}
 \clearfield{date}
 \clearfield{eprint}
 \clearfield{isbn}
 \clearfield{issn}
 \clearlist{location}
 \clearfield{month}
 \clearfield{day}
 \clearfield{series}
 \clearfield{volume}
 \clearfield{number}
 \clearfield{pages}
 \clearfield{note}
 \ifentrytype{book}{}{% Remove publisher and editor except for books
  \clearlist{publisher}
  \clearname{editor}
 }
}
\AtBeginBibliography{\footnotesize}
\begin{document}

%\title{A Comprehensive Exploration of Individual Human Brain Structure-Function Relationships using Graph Signal Processing}
\title{Coupled generator decomposition for fusion of electro- and magnetoencephalography data}

\author{Anders~S.~Olsen$^{1}$, 
Jesper~D.~Nielsen$^{2}$, 
Morten~Mørup$^{1*}$
\\
\\
\small $^{1}$\textit{Department of Applied Mathematics and Computer Science, Technical University of Denmark, Kgs. Lyngby, Denmark}
\\
\small $^{2}$\textit{Danish Research Centre for Magnetic Resonance, Centre for Functional and Diagnostic Imaging and Research,}\\\textit{Copenhagen University Hospital Amager and Hvidovre, Denmark}
\\
* \textit{Corresponding author}}
\renewcommand{\arraystretch}{1.2}

\maketitle

\begin{abstract}

Data fusion modeling can identify common features across diverse data sources while accounting for source-specific variability. Here we introduce the concept of a \textit{coupled generator decomposition} 
and demonstrate how it generalizes sparse principal component analysis (SPCA) for data fusion. Leveraging data from a multisubject, multimodal (electro- and magnetoencephalography (EEG and MEG)) neuroimaging experiment, we demonstrate the efficacy of the framework in identifying common features in response to face perception stimuli, while accommodating modality- and subject-specific variability. Through split-half cross-validation of EEG/MEG trials, we investigate the optimal model order and regularization strengths for models of varying complexity, comparing these to a group-level model assuming shared brain responses to stimuli. Our findings reveal altered $\sim170ms$ fusiform face area activation for scrambled faces, as opposed to real faces, particularly evident in the multimodal, multisubject model. Model parameters were inferred using stochastic optimization in PyTorch, demonstrating comparable performance to conventional quadratic programming inference for SPCA but with considerably faster execution. We provide an easily accessible toolbox for coupled generator decomposition that includes data fusion for SPCA, archetypal analysis and directional archetypal analysis. Overall, our approach offers a promising new avenue for data fusion.
\end{abstract}


\section{Introduction}

Recent advances in neuroimaging techniques have enabled researchers to concurrently collect data from multiple modalities and subjects, offering new opportunities for understanding the neural underpinnings of cognitive processes. Simultaneous measurements of electroencephalography (EEG), capturing electrical potentials on the scalp, and magnetoencephalography (MEG), measuring magnetic field strength offer a prime example. While both modalities detect synchronized postsynaptic activity in the dendrites of cortical pyramidal neurons, they differ in their sensitivity to source depth, tissue-specific signal attenuation, and source orientation, making their combination valuable for understanding signal sources and intermodal differences \cite{baillet_magnetoencephalography_2017,lopes_da_silva_eeg_2013,ahlfors_meg_2019}.

Existing approaches to M/EEG fusion include modality-specific error weighting using Bayesian optimization \cite{henson_meg_2009}, modality dissimilarity correlation modeling \cite{cichy_similarity-based_2016}, and fusion for source estimation \cite{baillet_electromagnetic_2001, chowdhury_megeeg_2015}, More recently, archetypal analysis (AA) \cite{cutler_archetypal_1994,morup_archetypal_2012}, which locates extremes or corners in the data and reconstructs data as convex combinations of these points, has been extended to multisubject modeling \cite{hinrich_archetypal_2016}, particularly relevant for M/EEG microstate modeling. Building on this, we proposed a polarity and scale-invariant multimodal and multisubject version \cite{olsen_combining_2022}. 

In this paper, we further develop the concept of a \textit{coupled generator decomposition}, where shared features are uncovered across data sources, while allowing to account for data source-specific spatiotemporal variability. Within the coupled generator decomposition we propose a multimodal, multisubject extension to sparse principal component analysis (SPCA) to uncover source-specific features while preserving component correspondence. Whereas SPCA has been considered previously in the context of functional neuroimaging \cite{sjostrand2006sparse}, to the best of our knowledge, we are the first to consider SPCA in the data fusion setting. We showcase the model on evoked response potential (ERP) data of face perception stimuli across the two neuroimaging modalities and subjects. Our framework assumes that at least one dimension is shared across entities. In this case, the time axis is shared and thus, the timing of the neural response to stimuli is assumed equal across modalities and subjects. Contrarily, the number of sensors may differ. SPCA is a well-established dimensionality reduction technique, which seeks to identify a small number of "sparse" principal components, highlighting time points in the ERP deemed most important for each component. The product of the generator matrix with the original data reveals modality- and subject-specific topographic maps such that the original data is reconstructed through modality- and subject-specific time-courses (i.e., mixing matrices), which reveal the trajectory taken in the evoked response. 

%One cognitive process that has received considerable attention is face perception, a complex, dynamic process that involves multiple sequentially activated brain regions. A popular data set acquired simultaneous EEG and MEG on 19 subjects conducting an experiment in which they viewed approximately 150 images of famous, unfamiliar, or scrambled faces \cite{wakeman_multi-subject_2015}. Previous analyses have revealed a positive  negative 170ms component corresponding to fusiform gyrus activation \cite{gao_neural_2019}, which was significantly larger for the non-scrambled conditions \cite{wakeman_multi-subject_2015} and displayed a different topography \cite{olsen_combining_2022}, and a ~400ms frontoparietal component exclusive for the face conditions though stronger for the familiar faces opposing the unfamiliar faces. 

Importantly, we develop our models in the PyTorch optimization framework, enabling gradient-based likelihood optimization through automatic differentiation, thereby alleviating the analytical derivation of model gradients and model-specific tailored optimization procedures. We compare the sum-of-squared-errors (SSE) performance to the more conventional quadratic programming algorithm for SPCA \cite{zou_sparse_2006}. We select the optimal model order and regularization parameters via split-half cross-validation, and compare error evolution across model orders with a group SPCA formulation and multimodal multisubject AA \cite{hinrich_archetypal_2016,olsen_combining_2022}. Overall, our approach provides a promising new avenue for understanding shared neural features in complex cognitive processes, with potential applications beyond neuroscience research. %By addressing the challenges posed by the combination of EEG and MEG data, we hope to provide a more complete understanding of the neural basis of cognitive processes such as face perception, and to identify new avenues for future research.

\section{Methods}

\begin{figure}
    \centering
    \includegraphics[width=\columnwidth]{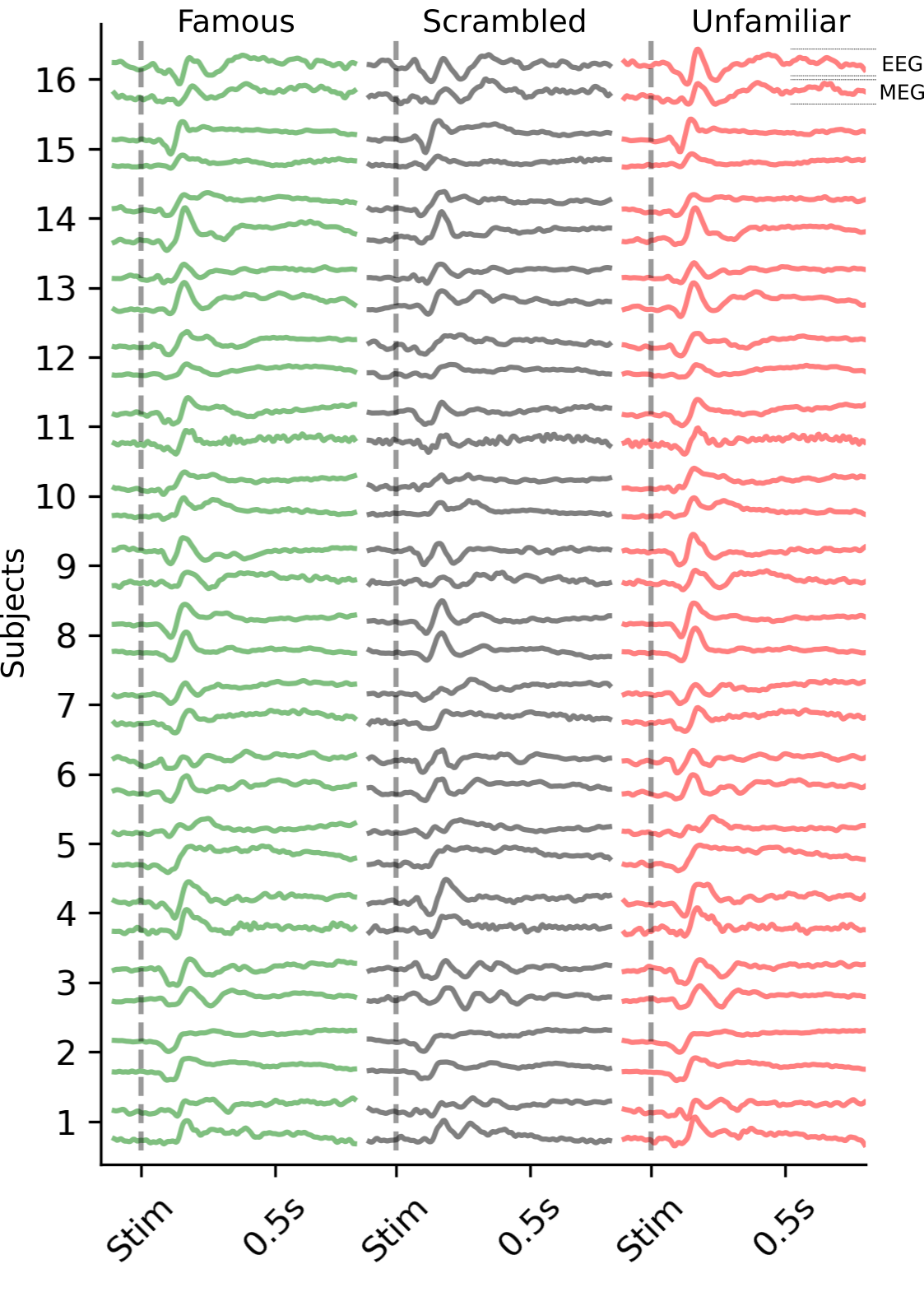}
    \caption{Variability in ERP waveform across subjects for a chosen right occipital EEG and MEG channel.}
    \label{fig:example}
\end{figure}
\subsection{Coupled generator decomposition}

We define a general linear matrix decomposition framework that minimizes the sum-of-squared-errors (SSE) reconstruction loss, thereby assuming a Gaussian noise structure:

\begin{equation}\label{eq:matrixdecom}
    \argmin_{\G,\S} ||\X-\X\G\S||_F^2
\end{equation}

The matrix $\G\in\mathbb{R}^{P\times K}$, denoted the generator matrix, linearly projects the $P$-dimensional data $\X\in\mathbb{R}^{N\times P}$ into a $K$-dimensional latent (source) space, with reconstruction facilitated by $\S\in\mathbb{R}^{K\times P}$. The solution to Eq.~\ref{eq:matrixdecom} requires constraints to yield meaningful latent structures. Principal component analysis (PCA) imposes an orthogonality constraint in the columns of $\S$. Such a solution is not unique, though, due to the rotational invariance of PCA, and additional constraints are needed to ensure uniqueness promoting interpretable results. For example, independent component analysis (ICA) assumes statistical independence in the sources $\X\G$, while archetypal analysis introduces non-negativity and sum-to-one constraints on the columns of $\G$ and $\S$. 

The data may be composed of several matrices $\X^{(b)}, b=\{1,\ldots,B\}$ representing, e.g., $B$ subjects in a multisubject experiment. The SSE equation then becomes 

\begin{equation}
    \argmin_{\G,\S^{(b)}} \sum_{b=1}^B||\X^{(b)}-\X^{(b)}\G\S^{(b)}||_F^2.
\end{equation}

In this context, the generator matrix $\G$ is shared across subjects while sources $\X^{(b)}\G$ and mixing matrices $\S^{(b)}$ are data matrix specific. This framework only requires that $P$ is a shared dimension across subjects while $N$ may differ, i.e., $\X^{(b)}\in\mathbb{R}^{N_b\times P}$. 

The framework may be further extended to also include multiple modalities $m$, i.e., $\mathbf{X}^{(m,b)}$. Then $\G$ captures the information that is shared across the entire data set, while $\mathbf{S}^{(m,b)}$ contains the reconstruction information for the specific individual and modality. Previously, archetypal analysis has been used for multisubject fMRI data, where $P$ represented voxels and $N$ the time axis of potentially differing length, such that neural activity source location was assumed equal across subjects \cite{hinrich_archetypal_2016}. In another example, $P$ was the time axis and $N$ the (differing) number of sensors across EEG and MEG such that the timing of neural response to stimuli was assumed equal across subjects and modalities \cite{olsen_combining_2022}. We omit the subject-and modality-specific notation in the following. 

We note that data reconstruction may be performed with an altered version of the data, $\Tilde{\X}$: 

\begin{equation}
    \argmin_{\G,\S} ||\X-\Tilde{\X}\G\S||_F^2,
\end{equation} 
a necessity for directional archetypal analysis \cite{olsen_combining_2022}, where the data is assumed $l_2$-normalized over the dimension $P$, i.e., $||\x_p||_2=1$, but the strength of the reconstruction driven by data points $\x_p$ of high amplitude. 

\subsection{Sparse principal component analysis}

Sparse PCA introduces uniqueness to Eq.~\ref{eq:matrixdecom} by adding  $l_1$- and $l_2$-regularization terms on $\G$:

\begin{equation}\label{eq:mmmsSPCA}
    \argmin_{\G,\S} ||\X-\Tilde{\X}\G\S||_F^2 + \lambda_2 \sum_{k=1}^K ||\G_k||^2 + \lambda_1 \sum_{k=1}^K ||\G_k||_1
\end{equation}

subject to $\S^\top\S=\I$. A quadratic programming algorithm to minimize Eq.~\ref{eq:mmmsSPCA} originally proposed by \cite{zou_sparse_2006} alternates between updating $\G$ and $\S$. The source matrix $\G$ is updated using an elastic net estimate, while $\S$ is given by a Procrustes rotation through the singular value decomposition (SVD) to ensure orthogonal columns in $\S$:

\begin{align}\label{eq:Supdate}
    (\X^\top\X)\G&=\mathbf{U}\boldsymbol{\Sigma}\mathbf{V}^\top\\
    \mathbf{S}^\top & = \mathbf{U}\mathbf{V}^\top. \nonumber
\end{align}

\subsection{Computational implementation}

We leverage the PyTorch stochastic optimization framework and the Adam optimizer \cite{kingma_adam_2014} to minimize Eq.~\ref{eq:mmmsSPCA}. PyTorch has an efficient automatic numerical gradient computational structure, which only requires specifying a loss function and the parameters to be learned. To avoid gradient explosions due to the non-differentiable $l_1$-regularization term in SPCA, we introduce two non-negative matrices $\G=\G_p-\G_n$. The non-negativity constraint is implemented by optimizing the two matrices unconstrained and running them through a softplus function $x=\log(1+e^{x})$ before calculating the SSE. The mixing matrix $\S$ is inferred using Eq.~\ref{eq:Supdate} by propagating gradients in PyTorch through the SVD of $(\mathbf{X}^\top\mathbf{X})\G$, looping over subjects and modalities in the data fusion case, such that the optimization is reduced to only optimizing for the generator $\G$ accounting for $\S$ through its dependence on $\G$. For archetypal analysis, both $\G$ and $\S$ are learned, and the non-negativity and sum-to-one constraints are implemented through the softmax function. Though archetypal analysis uses the same loss function as SPCA (without regularization terms), the directional variant is different and may be found in \cite{olsen_combining_2022}. We used a learning rate of $0.01$. Gradient optimization through non-full-rank SVD tends to be unstable, and thus, we measured relative convergence between the lowest and second-lowest loss over the latter 5 iterations, stopping when this relative convergence reached $10^{-8}$. 

Since the direct optimization of $\G$ accounting for its dependency on $\S$ can in general be prone to local minima due to the non-convex optimization problem, we implemented annealing to help escape some of the local minima. Here we implemented models for a range of $l_2$-regularization constants and for each value of $\lambda_2=\{10^{-5},\ldots,10^{0}\}$, we trained a model with $\lambda_1=0$, then trained a model initialized in this obtained solution using $\lambda_1=10^{-5}$ and continued this procedure using the previous model as a starting point for the ranges of $\lambda_1=\{10^{-5},\ldots,10^{0}\}$. 

The Python-implementation is available as a toolbox here: \href{https://github.com/anders-s-olsen/Coupled-generator-decomposition}{github.com/anders-s-olsen/Coupled-generator-decomposition}. 

\subsection{Experimental data and preprocessing}

We used the openly available multimodal face perception dataset with simultaneously recorded 70-channel electro- and 102-channel magnetoencephalography data for 19 subjects, of which three were excluded by the original authors due to poor data quality\footnote{\url{https://openneuro.org/datasets/ds000117/versions/1.0.4}} \cite{wakeman_multi-subject_2015}. Each participant were presented approximately 900 images of famous, unfamiliar, or scrambled faces. Data were preprocessed using MNE \cite{gramfort_meg_2013}. Artifact trials were removed and the remaining trials randomly distributed into a train and a test set. Data were bandpass filtered between 0.5--40 Hz and subsequently demeaned channel-wise, and downsampled from 1100 Hz to 200 Hz. Trials were randomly split into a training and test set, and trials in these were averaged into ERPs within-subject, within-group, within-condition. The test set was subsequently further split into a validation and a test set constituting the first and latter 8 subjects, respectively. Validation loss was evaluated using $SSE_{val}=||\X_{test}^{(m,b)}-\X_{train}^{(m,b)}\G\S^{(m,b)}||$, where $\G$ and $\S^{(m,b)}$ were learned from the training data, similarly for test loss. The training data for a selected right-occipital EEG and MEG channel may be seen in Fig.~\ref{fig:example}. 

We investigated this data in 1) group, 2) multimodal, and 3) multimodal, multisubject formulations by concatenating data in different ways. For $B=16$ subjects, $M=2$ modalities, $P=180$ time points from $-100ms$ to $800ms$ and $N_m\in\{70,102\}$ channels for EEG and MEG, respectively, and three conditions (famous, scrambled, unfamiliar), the group analysis was carried out on ERPs stacked vertically across subjects and modalities, i.e., $\X_{group}\in\mathbb{R}^{BN_{EEG}+BN_{MEG}\times 3P}$. Multimodal analysis was carried out on $\X_{mm}^{(m)}\in\mathbb{R}^{BN_m\times 3P}$, where the loss function and SVD was computed as a loop over $m$, and similarly for multimodal, multisubject: $\X_{mmms}^{(m,b)}\in\mathbb{R}^{N_m\times 3P}$. The three ERPs corresponding to different face perception stimuli were stacked temporally to allow for condition-specific trajectories while component topographies were assumed equal, as suggested in a previous study \cite{olsen_combining_2022}.

\section{Results and Discussion}
\label{sec:results} 

\begin{figure}
    \centering
    \includegraphics[width=\columnwidth]{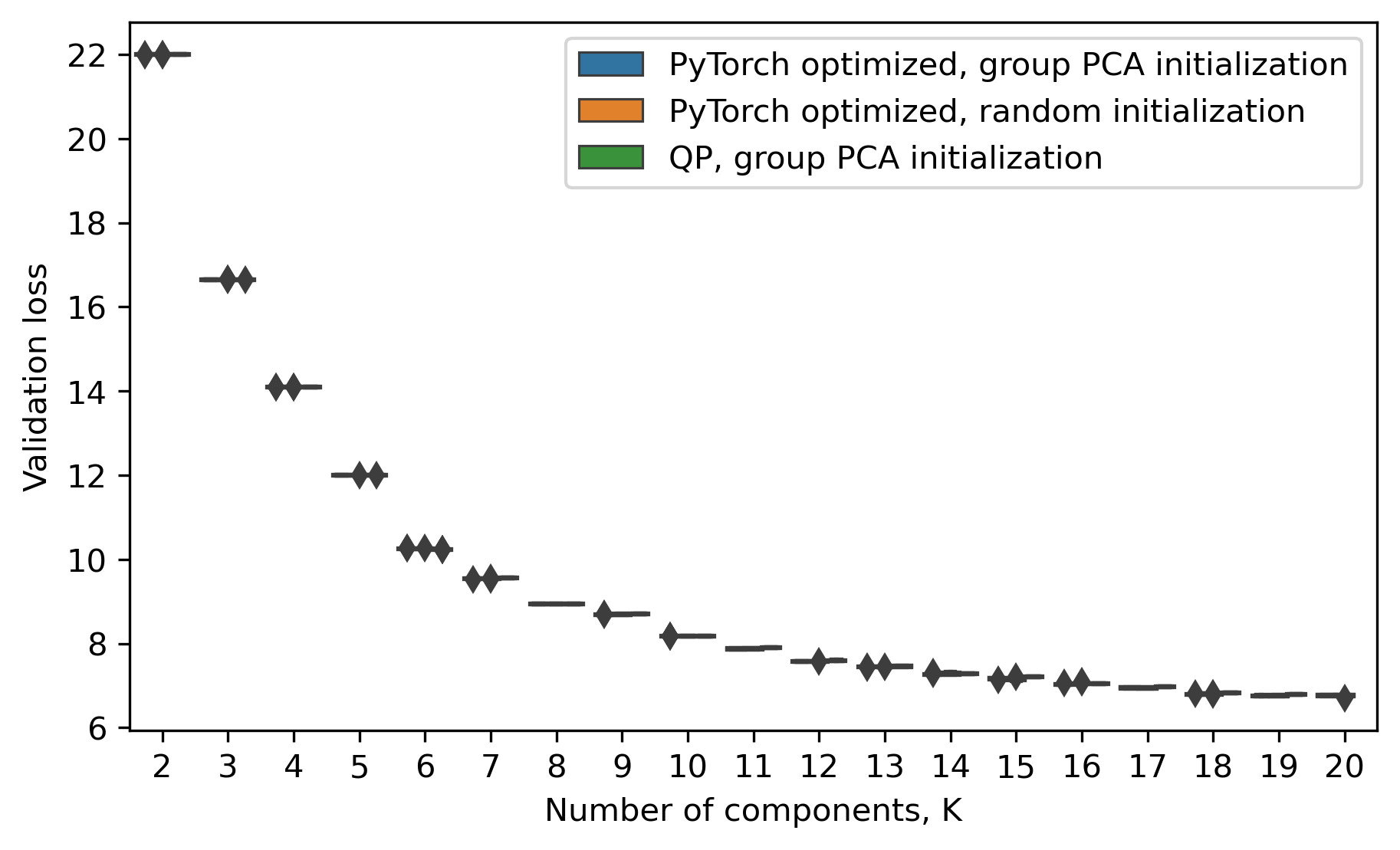}
    \caption{Boxplot of model convergence across stochastic optimization in PyTorch and quadratic programming inference techniques, as well as effect of initialization on validation loss. Each model was run across 10 random initializations using a regularization coefficient pair determined on the lowest attained validation loss. We note that there are, in fact, three boxes for each number of components; the boxes are very small due to diminishing variability. }
    \label{fig:model_loss}
\end{figure}

To evaluate the effectiveness of the stochastic gradient optimization framework in PyTorch against traditional quadratic programming inference for sparse PCA, we computed group sparse PCA formulations for varying number of components $K\in\{2,\ldots 20\}$. Figure \ref{fig:model_loss} displays the convergence of the models across 10 random initializations. Both models were initialized using a rank-$K$ group PCA,, i.e., $\G=\mathbf V_K$, where $\mathbf V_K$ were the first $K$ columns of the right principal components of $\X_{group}$. We also compare these results to the PyTorch model, where $\G$ was initialized using a standard normal distribution.  Notably, the three formulations exhibited comparable performance and low variability across initializations. The computational time was drastically lower for the PyTorch optimization than QP (minutes vs hours CPU-time for high $K$), and group-PCA initialization converged the fastest. These results establish PyTorch as the preferred choice for its ease of implementation and faster convergence. 

\begin{table}[]\label{tab:tab}
\caption{The regularization parameters leading to the lowest validation loss for different model orders. MMMS = Multimodal, multisubject.}
\centering
\begin{tabular}{c|cc|cc|cc|}
\cline{2-7}
\multirow{2}{*}{K} & \multicolumn{2}{c|}{Group} & \multicolumn{2}{c|}{Multimodal}   & \multicolumn{2}{c|}{MMMS} \\ \cline{2-7} 
& $\lambda_1$  & $\lambda_2$ & $\lambda_1$ & $\lambda_2$ & $\lambda_1$ & $\lambda_2$ \\ \cline{2-7} 
2 & $10^{-4}$ & $10^{-4}$ & $10^{-4}$ & $10^{-2}$ & $10^{-5}$ & $10^{-4}$ \\
3 & $10^{-3}$ & $10^{-1}$ & $10^{-4}$ & $0$ & $10^{-4}$ & $10^{-4}$ \\
4 & $10^{-5}$ & $10^{-1}$ & $0$ & $0$ & $10^{-2}$ & $10^{-5}$ \\
5 & $10^{-3}$ & $10^{-1}$ & $10^{-5}$ & $0$ & $10^{-2}$ & $10^{-1}$ \\
6 & $10^{-2}$ & $10^{-2}$ & $10^{-2}$ & $10^{-1}$ & $10^{-2}$ & $10^{-1}$ \\
7 & $10^{-2}$ & $10^{-1}$ & $10^{-2}$ & $10^{-1}$ & $10^{-2}$ & $10^{-1}$ \\
8 & $10^{-2}$ & $10^{-1}$ & $10^{-2}$ & $10^{-1}$ & $10^{-2}$ & $10^{-1}$ \\
9 & $10^{-2}$ & $10^{-1}$ & $10^{-2}$ & $10^{-1}$ & $10^{-2}$ & $10^{-1}$ \\
10 & $10^{-2}$ & $10^{-1}$ & $10^{-3}$ & $10^{-1}$ & $10^{-2}$ & $10^{-1}$ \\
11 & $10^{-2}$ & $10^{-1}$ & $10^{-2}$ & $10^{-1}$ & $10^{-1}$ & $10^{-1}$ \\
12 & $10^{-2}$ & $10^{-1}$ & $10^{-2}$ & $10^{-1}$ & $10^{-1}$ & $10^{-1}$ \\
13 & $10^{-2}$ & $10^{-1}$ & $10^{-2}$ & $10^{-1}$ & $10^{-1}$ & $10^{-1}$ \\
14 & $10^{-2}$ & $10^{-1}$ & $10^{-2}$ & $10^{-1}$ & $10^{-1}$ & $10^{-1}$ \\
15 & $10^{-2}$ & $10^{-1}$ & $10^{-2}$ & $10^{-1}$ & $10^{-2}$ & $10^{-1}$ \\
16 & $10^{-2}$ & $10^{-1}$ & $10^{-2}$ & $10^{-1}$ & $10^{-2}$ & $10^{-1}$ \\
17 & $10^{-2}$ & $10^{-1}$ & $10^{-2}$ & $10^{-1}$ & $10^{-2}$ & $10^{-1}$ \\
18 & $10^{-2}$ & $10^{-1}$ & $10^{-2}$ & $10^{-1}$ & $10^{-2}$ & $10^{-1}$ \\
19 & $10^{-2}$ & $10^{-1}$ & $10^{-2}$ & $10^{-1}$ & $10^{-2}$ & $10^{-1}$ \\
20 & $10^{-2}$ & $10^{-1}$ & $10^{-2}$ & $10^{-1}$ & $10^{-2}$ & $10^{-1}$ \\ \cline{2-7}
\end{tabular}
\end{table}

Next, we explored the optimal regularization coefficients for PyTorch optimized, group-PCA initialized group, multimodal, and multimodal, multisubject sparse PCA (see Tab.~1). Regularization coefficients were selected based on the lowest attained validation loss. Higher regularization led to improved performance upon increasing model order (number of components). Interestingly, the need for regularization became evident at lower model orders for the more complex multimodal, multisubject model, suggesting a sensitivity to overfitting. However, from $K=6$ and onwards, all models agreed on $\lambda_2=10^{-1}$ and sparsity coefficient $\lambda_1=10^{-1}$ or $\lambda_1=10^{-2}$. Thus, for sufficient model complexity, there is a clear advantage of both types of regularization. 

\begin{figure}[h]
    \centering
    \includegraphics[width=\columnwidth]{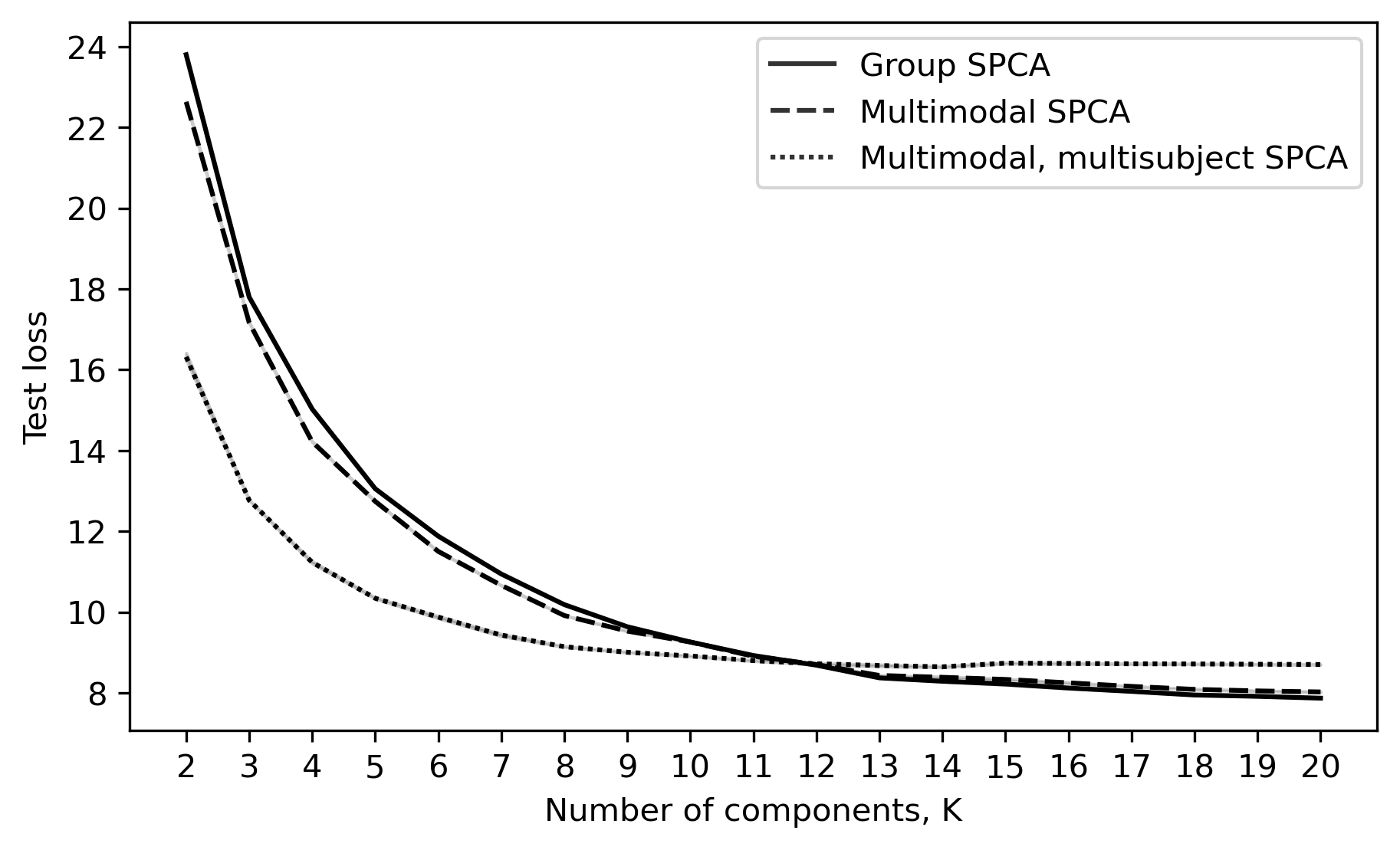}
    \caption{Lineplot of model performance across number of components, $K$. The models were evaluated on a test set using regularization coefficients determined on the validation set, and the average and standard deviation (shaded area) across 10 random initializations is shown. We note that the variance across random initializations is too small to be clearly distribguishable.}
    \label{fig:model_order}
\end{figure}

We investigated the test loss, examining the latter 8 subjects in the test set as a function of model order using the determined regularization coefficients (Fig.~\ref{fig:model_order}). We observed reduced test loss for multimodal, multisubject SPCA for low model orders, emphasizing its expressive power. However, when model order increased, the gain of using the expressive formulation decreased and the group SPCA formulation attained the lowest test loss for high number of components. The crossover happened at $K=12$. Group SPCA and multimodal SPCA were again very similar. Notably, no distinct "elbow" in the loss curve was observed, but a midway-point for the multimodal, multisubject version is suggested at $K=5$, where the gain in test loss diminished with additional components. 

\begin{figure*}
    \centering
    \includegraphics[width=\textwidth]{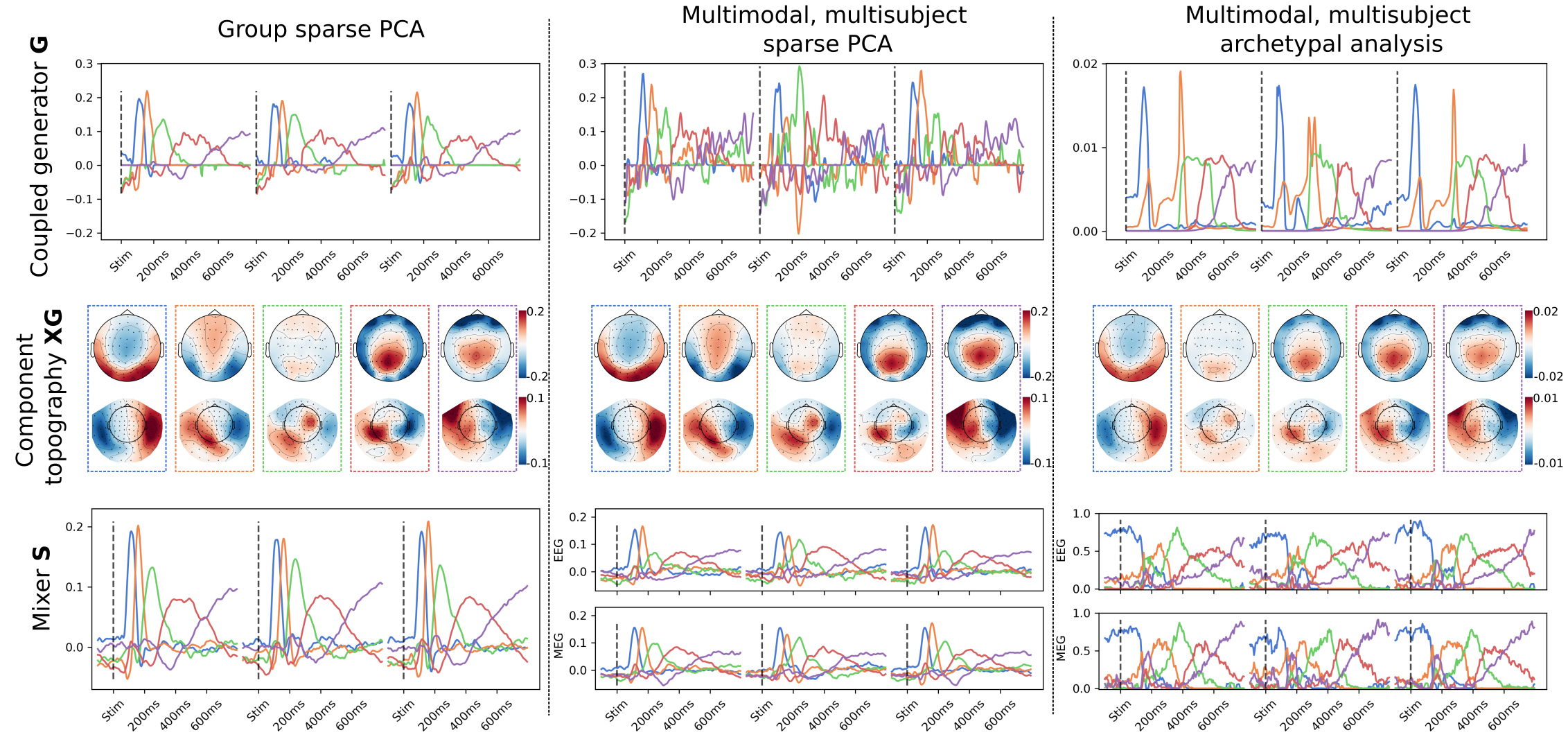}
    \caption{Sparse PCA and archetypal analysis results on data from a multimodal multisubject face perception neuroimaging experiment. The coupled generator decomposition computes a shared generator matrix $\G$ (top row) and subject and modality-specific topographical maps $\X\G$, here shown as an average across subjects. For the group formulation, the mixing matrix $\S$ is also shared, while we show the average mixing matrix across subjects for the multimodal, multisubject models. EEG units are in $\mu V$ and MEG units are in $fT$.}
    \label{fig:results}
\end{figure*}

Fig.~\ref{fig:results} presents the parameters of the group and multimodal, multisubject SPCA for $K=5$ components for optimal regularization parameters $\lambda_1=10^{-2}$ and $\lambda_2=10^{-1}$. Here, $\Tilde{\X}$ was only the poststimulus part of the data, and thus, the generator matrix $\G$ only contains the post-stimulus part of the data, while the mixing matrices $\S$ include the prestimulus part. 

While the generator matrix $\G$ was more smooth for the group formulation, the order of components and their topography (shown as an average across subjects) remained consistent: 1) a $\sim\!100ms$ component with strong occipital EEG activation and left-right MEG symmetry, 2) a $\sim\!170ms$ component with lateral occipital vs frontal topography, 3) a $\sim\!230ms$ frontoparietal vs lateral occipital component with weak topography, 4) a $\sim\!350-550ms$ parietal component, and lastly 5) a frontal component dominant from $550ms$ and onwards. The $\sim\!170ms$ component was less strong for the scrambled condition and even contained a negative $\sim\!230ms$ deflection in the multimodal, multisubject version. This altered $\sim\!170ms$ deflection for scrambled faces has also been observed previously \cite{wakeman_multi-subject_2015,olsen_combining_2022} and falls in line with the observation that this component corresponds to fusiform face area activation \cite{gao_neural_2019}. The mixing matrices, which for the multimodal, multisubject version is shown as an average across subjects, were generally smooth and similar across models and modalities while still displaying the altered $\sim\!170ms$ (orange) component for scrambled faces. The subject-specific mixing matrices are shown in the readme of \href{https://github.com/anders-s-olsen/Coupled-generator-decomposition}{our github repository} and generally show high subject variability. 

We compare these results with a multimodal, multisubject archetypal analysis also implemented in the same toolbox and optimized using stochastic optimization in PyTorch. Archetypal analysis naturally learns a very sparse generator matrix $\G$ due to the non-negativity and sum-to-one constraints. Thus, sources are pinpointed more precisely. Here, the lateral occipital vs frontal $\sim\!170ms$ component has disappeared completely, while the parietal component was split in two components with similar topography, of which the first one of the two ($\sim\!200-300ms$) is activated slightly more strongly for the famous and unfamiliar faces than in the scrambled condition. Combined, these results suggest that the difference between the three face conditions is small but most notable in the $\sim\!170ms$ component. 

\section{Conclusions}
Here we presented a unified data fusion framework, where data, comprising, e.g., multiple data sources such as subjects or modalities, which share at least one dimension, may be modeled together with a coupled generator matrix and source-specific mixing matrices. The framework is general and easily extended from sparse PCA to, e.g., Gaussian or directional archetypal analysis using different parameter constraints and loss functions, i.e., using only a couple lines of code. Notably, our stochastic optimization in PyTorch is fast while performing equally well to established inference techniques, i.e., quadratic programming for sparse PCA. 

%\bibliography{mm, references,references2}
\printbibliography

@inproceedings{sjostrand2006sparse,
  title={Sparse PCA, a new method for unsupervised analyses of fmri data},
  author={Sj{\"o}strand, Karl and Lund, Torben Ellegaard and Madsen, Kristoffer Hougaard and Larsen, Rasmus},
  booktitle={Proc. International Society of Magnetic Resonance In Medicine-ISMRM},
  pages={1--1},
  year={2006}
}

@article{gramfort_meg_2013,
	title = {{MEG} and {EEG} data analysis with {MNE}-Python},
	volume = {7},
	issn = {1662-453X},
	url = {https://www.frontiersin.org/articles/10.3389/fnins.2013.00267},
	journaltitle = {Frontiers in Neuroscience},
	author = {Gramfort, Alexandre and Luessi, Martin and Larson, Eric and Engemann, Denis and Strohmeier, Daniel and Brodbeck, Christian and Goj, Roman and Jas, Mainak and Brooks, Teon and Parkkonen, Lauri and Hämäläinen, Matti},
	urldate = {2024-01-30},
	date = {2013},
}

@article{zou_sparse_2006,
	title = {Sparse Principal Component Analysis},
	volume = {15},
	issn = {1061-8600},
	url = {https://www.jstor.org/stable/27594179},
	pages = {265--286},
	number = {2},
	journaltitle = {Journal of Computational and Graphical Statistics},
	author = {Zou, Hui and Hastie, Trevor and Tibshirani, Robert},
	urldate = {2024-01-30},
	date = {2006},
	note = {Publisher: [American Statistical Association, Taylor \& Francis, Ltd., Institute of Mathematical Statistics, Interface Foundation of America]},
}

@incollection{ahlfors_meg_2019,
	location = {Cham},
	title = {{MEG} and Multimodal Integration},
	isbn = {978-3-319-62657-4},
	url = {https://doi.org/10.1007/978-3-319-62657-4_7-1},
	pages = {1--20},
	booktitle = {Magnetoencephalography: From Signals to Dynamic Cortical Networks},
	publisher = {Springer International Publishing},
	author = {Ahlfors, Seppo P.},
	editor = {Supek, Selma and Aine, Cheryl J.},
	urldate = {2024-01-30},
	date = {2019},
	langid = {english},
	doi = {10.1007/978-3-319-62657-4_7-1},
}

@article{baillet_electromagnetic_2001,
	title = {Electromagnetic brain mapping},
	volume = {18},
	issn = {1558-0792},
	doi = {10.1109/79.962275},
	pages = {14--30},
	number = {6},
	journaltitle = {{IEEE} Signal Processing Magazine},
	author = {Baillet, S. and Mosher, J.C. and Leahy, R.M.},
	date = {2001-11},
	note = {Conference Name: {IEEE} Signal Processing Magazine},
}

@article{olsen_combining_2022,
	title = {Combining electro- and magnetoencephalography data using directional archetypal analysis},
	volume = {16},
	issn = {1662-453X},
	url = {https://www.frontiersin.org/articles/10.3389/fnins.2022.911034},
	journaltitle = {Frontiers in Neuroscience},
	author = {Olsen, Anders S. and Høegh, Rasmus M. T. and Hinrich, Jesper L. and Madsen, Kristoffer H. and Mørup, Morten},
	urldate = {2023-05-02},
	date = {2022},
}

@article{henson_meg_2009,
	title = {{MEG} and {EEG} data fusion: Simultaneous localisation of face-evoked responses},
	volume = {47},
	issn = {1053-8119},
	url = {https://www.ncbi.nlm.nih.gov/pmc/articles/PMC2912501/},
	doi = {10.1016/j.neuroimage.2009.04.063},
	shorttitle = {{MEG} and {EEG} data fusion},
	pages = {581--589},
	number = {2},
	journaltitle = {{NeuroImage}},
	shortjournal = {Neuroimage},
	author = {Henson, Richard N. and Mouchlianitis, Elias and Friston, Karl J.},
	urldate = {2023-05-02},
	date = {2009-08-15},
	pmid = {19398023},
	pmcid = {PMC2912501},
}

@article{baillet_magnetoencephalography_2017,
	title = {Magnetoencephalography for brain electrophysiology and imaging},
	volume = {20},
	rights = {2017 Nature Publishing Group, a division of Macmillan Publishers Limited. All Rights Reserved.},
	issn = {1546-1726},
	url = {https://www.nature.com/articles/nn.4504},
	doi = {10.1038/nn.4504},
	pages = {327--339},
	number = {3},
	journaltitle = {Nature Neuroscience},
	shortjournal = {Nat Neurosci},
	author = {Baillet, Sylvain},
	urldate = {2023-05-02},
	date = {2017-03},
	langid = {english},
	note = {Number: 3
Publisher: Nature Publishing Group},
}

@article{lopes_da_silva_eeg_2013,
	title = {{EEG} and {MEG}: Relevance to Neuroscience},
	volume = {80},
	issn = {0896-6273},
	doi = {10.1016/J.NEURON.2013.10.017},
	pages = {1112--1128},
	number = {5},
	journaltitle = {Neuron},
	author = {Lopes da Silva, Fernando},
	urldate = {2022-01-24},
	date = {2013-12-04},
	pmid = {24314724},
	note = {Publisher: Cell Press},
}

@article{morup_archetypal_2012,
	title = {Archetypal analysis for machine learning and data mining},
	volume = {80},
	issn = {0925-2312},
	url = {https://orbit.dtu.dk/en/publications/archetypal-analysis-for-machine-learning-and-data-mining},
	doi = {10.1016/J.NEUCOM.2011.06.033},
	pages = {54--63},
	journaltitle = {Neurocomputing},
	author = {Mørup, Morten and Hansen, Lars Kai},
	urldate = {2022-01-24},
	date = {2012-03-15},
	note = {Publisher: Elsevier},
}

@article{cutler_archetypal_1994,
	title = {Archetypal Analysis},
	volume = {36},
	issn = {00401706},
	doi = {10.2307/1269949},
	pages = {338},
	number = {4},
	journaltitle = {Technometrics},
	author = {Cutler, Adele and Breiman, Leo},
	urldate = {2022-01-24},
	date = {1994-11},
	note = {Publisher: {JSTOR}},
}

@article{wakeman_multi-subject_2015,
	title = {A multi-subject, multi-modal human neuroimaging dataset},
	volume = {2},
	issn = {2052-4463},
	url = {https://www.nature.com/articles/sdata20151},
	doi = {10.1038/sdata.2015.1},
	pages = {1--10},
	number = {1},
	journaltitle = {Scientific Data 2015 2:1},
	author = {Wakeman, Daniel G. and Henson, Richard N.},
	urldate = {2022-01-24},
	date = {2015-01-20},
	pmid = {25977808},
	note = {Publisher: Nature Publishing Group},
}

@article{kingma_adam_2014,
	title = {Adam: A Method for Stochastic Optimization},
	url = {https://arxiv.org/abs/1412.6980v9},
	doi = {10.48550/arxiv.1412.6980},
	journaltitle = {3rd International Conference on Learning Representations, {ICLR} 2015 - Conference Track Proceedings},
	author = {Kingma, Diederik P. and Ba, Jimmy Lei},
	urldate = {2023-03-02},
	date = {2014-12-22},
	eprinttype = {arxiv},
	eprint = {1412.6980},
	note = {Publisher: International Conference on Learning Representations, {ICLR}},
}

@article{gao_neural_2019,
	title = {The neural sources of N170: Understanding timing of activation in face-selective areas},
	volume = {56},
	issn = {14698986},
	doi = {10.1111/psyp.13336},
	number = {6},
	journaltitle = {Psychophysiology},
	author = {Gao, Chuanji and Conte, Stefania and Richards, John E. and Xie, Wanze and Hanayik, Taylor},
	date = {2019},
}

@article{chowdhury_megeeg_2015,
	title = {{MEG}–{EEG} Information Fusion and Electromagnetic Source Imaging: From Theory to Clinical Application in Epilepsy},
	volume = {28},
	issn = {15736792},
	doi = {10.1007/s10548-015-0437-3},
	number = {6},
	journaltitle = {Brain Topography},
	author = {Chowdhury, Rasheda Arman and Zerouali, Younes and Hedrich, Tanguy and Heers, Marcel and Kobayashi, Eliane and Lina, Jean Marc and Grova, Christophe},
	date = {2015},
}

@article{cichy_similarity-based_2016,
	title = {Similarity-Based Fusion of {MEG} and {fMRI} Reveals Spatio-Temporal Dynamics in Human Cortex During Visual Object Recognition},
	volume = {26},
	issn = {14602199},
	doi = {10.1093/cercor/bhw135},
	number = {8},
	journaltitle = {Cerebral Cortex},
	author = {Cichy, Radoslaw Martin and Pantazis, Dimitrios and Oliva, Aude},
	date = {2016},
}

@article{hinrich_archetypal_2016,
	title = {Archetypal Analysis for Modeling Multisubject {fMRI} Data},
	volume = {10},
	issn = {19324553},
	doi = {10.1109/JSTSP.2016.2595103},
	pages = {1160--1171},
	number = {7},
	journaltitle = {{IEEE} Journal on Selected Topics in Signal Processing},
	author = {Hinrich, Jesper Love and Bardenfleth, Sophia Elizabeth and Roge, Rasmus Erbou and Churchill, Nathan William and Madsen, Kristoffer Hougaard and Morup, Morten},
	urldate = {2022-01-24},
	date = {2016-10-01},
	note = {Publisher: Institute of Electrical and Electronics Engineers Inc.},
}
\end{document}